\documentclass[amssymb,amsmath,aps,showpacs,floatfix,nofootinbib,showpacs,12pt]{revtex4}
\usepackage{color}
\usepackage{graphicx}
\def\beq{\begin{equation}}
\def\eeq{\end{equation}}

\def\be{\begin{equation}}
\def\ee{\end{equation}}
\def\bea{\begin{eqnarray}}
\def\eea{\end{eqnarray}}

\newcommand{\gsim}{\lower.7ex\hbox{$\;\stackrel{\textstyle>}{\sim}\;$}}
\newcommand{\lsim}{\lower.7ex\hbox{$\;\stackrel{\textstyle<}{\sim}\;$}}

\begin{document}

\hspace{4.5in}{}

\bigskip

\title{Minimal Modification to Tri-bimaximal Mixing}

\author{Xiao-Gang He$^{1,2}$ and A. Zee$^3$}
\affiliation{
$^1$INPAC, Department of Physics, Shanghai Jiao Tong University, Shanghai\\
$^2$Department of Physics and Center for Theoretical Sciences, National Taiwan University, Taipei\\
$^3$Kavli Institute for Theoretical Physics and Department of Physics, UCSB, Santa Barbara, CA 93106}

\begin{abstract}
We explore some ways of minimally modifying the neutrino mixing matrix from tribimaximal, characterized by introducing at most one mixing angle and a CP violating phase thus extending our earlier work. One minimal modification, motivated to some extent by group theoretic considerations, is a simple case with the elements $V_{\alpha 2}$ of the second column in the mixing matrix equal to  $1/\sqrt{3}$. Modifications by keeping one of the columns or one of the rows unchanged from tri-bimaximal mixing all belong to the class of minimal modification. Some of the  cases have interesting experimentally testable consequences. In particular, the T2K and MINOS collaborations have recently reported indications of a non-zero $\theta_{13}$. For the cases we consider, the new data sharply constrain the
CP violating phase angle $\delta$, with $\delta$ close to 0 (in some cases) and $\pi$ disfavored.
\end{abstract}


\maketitle

\section{Introduction}

Mixing of different neutrino species has been established by various experiments~\cite{pdg}. Recently the T2K~\cite{t2k} and MINOS~\cite{minos-new} collaborations have reported indications of a non-zero $\theta_{13}$ providing new evidence for neutrino mixing and new information about mixing pattern.
The mixing can be represented by the Pontecorvo-Maki-Nakagawa-Sakata (PMNS)~\cite{pmns} mixing matrix $V$ in the charged current interaction of the $W$ boson with left
handed charged leptons $l_L$ and neutrinos $\nu_L$,
${\cal L} = - (g/\sqrt{2}) \bar l_L \gamma^\mu V \nu_L W^+_\mu + h.c.$.
The elements of the unitary matrix $V$ are usually indicated by $V_{\alpha j}$ with $\alpha= e,\mu,\tau,...$ and $j = 1,2,3,...$. With three neutrinos, there are
three mixing angles, one Dirac phase, and possibly two Majorana phases if neutrinos are Majorana particles.

It is possible to fit data from various experiments~\cite{data}, except for possible anomalies in the LSND,  MiniBoon~\cite{lsnd}, and MINOS~\cite{minos} data.
The pre-T2K data, with modified Gallium cross-section used by SAGE collaboration~\cite{sage}, on mixing and mass parameters can be summarized at 1$\sigma$(3$\sigma$) level as~\cite{data, hint},
\begin{eqnarray}
&&\Delta m^2_{21} = 7.59\pm 0.20(^{+0.61}_{-0.69}) \times 10^{-5} \mbox{eV}^2\;,\nonumber\\
&&\Delta m^2_{31} = -2.36\pm 0.11(\pm0.37) [+2.46\pm0.12(\pm0.37)] \times 10^{-3} \mbox{eV}^2\;,\nonumber\\
&&\theta_{12} = 34.5\pm 1.0(^{+3.2}_{-2.8})^\circ\;,\;\;\theta_{23} = 42.8^{+4.7}_{-2.9}(^{+10.7}_{-7.3})^\circ\;,\;\;\theta_{13} = 5.1^{+3.0}_{-3.3}(\le 12.0)^\circ\;.
\label{fit-pret2k}
\end{eqnarray}
Here the mixing parameters are those of the Particle Data Group (PDG) parametrization~\cite{pdg} (with $\theta_{12} \equiv \theta_{e2},\;\;\theta_{23} \equiv \theta_{\mu 3},\;\;\theta_{13}\equiv \theta_{e3}$). There is no direct experimental information on the phases $\delta_{PDG}$.


The mixing pattern is well approximated by the
so-called tri-bimaximal mixing pattern of the form~\cite{tri}
\begin{eqnarray}
V_{TB} = \left ( \begin{array}{rrr}
\sqrt{2\over 3}&{1\over \sqrt{3}}&0\\
-{1\over \sqrt{6}}&{1\over \sqrt{3}}&{1\over \sqrt{2}}\\
-{1\over \sqrt{6}}&{1\over \sqrt{3}}&-{1\over \sqrt{2}}
\end{array}
\right )\;. \label{vtri}
\end{eqnarray}
Note that with this mixing pattern, $\theta_{13}$ is equal to zero.

The recent T2K data on $\theta_{13}$ (and therefore $|V_{e3}|$) show that~\cite{t2k} at 90\% C.L. it is not zero with $\sin^22\theta_{13}$ in the range of $0.03(0.04) \sim 0.28(0.34)$ for normal (inverted) neutrino mass hierarchy. Data from MINOS\cite{minos-new} also disfavor $\theta_{13} =0$ at the 89\% C.L. level. This information provide an important constraint on theoretical model buildings for neutrino mixing~\cite{new}. Combining the recent T2K and MINOS data with previous neutrino oscillation data and using the new reactor flux in ref.\cite{reactor}, more stringent constraints on the mixing parameters than those given in eq.(\ref{fit-pret2k}) have been obtained\cite{fogli}. These new constraints are shown in Table \ref{fit}. We will use them for our later discussions.

\begin{table}[tbp]
 \begin{tabular}{|c|c|c|c|}\hline
 Parameter&$sin^2\theta_{12}$&$sin^2\theta_{23}$&$sin^2\theta_{13}$\\
 \hline
 Best fit&0.312 &0.42&0.025\\\hline
 1$\sigma$ range&0.296 - 0.329&0.39 - 0.50&0.018 - 0.032\\\hline
 2$\sigma$ range&0.280 - 0.347&0.36 - 0.60&0.012 - 0.041
   \\ \hline
   3$\sigma$ range&0.265 - 0.364& 0.34 - 0.64& 0.005 - 0.050\\ \hline
 \end{tabular}
\caption{Ranges for mixing parameters obtained in Ref.\cite{fogli}.}
 \label{fit}
\end{table}

The combined data show that $\theta_{13}$ is non-zero at 3$\sigma$ level. In fact the pre-T2K data already provides a hint that $\theta_{13}$ is non-zero at a more-than-1$\sigma$ level~\cite{hint}. A non-zero $V_{e3}$ would rule out tri-bimaximal mixing. Now the tri-bimaximal mixing prediction for the angle $\theta_{12}$ is outside the 1$\sigma$. Also, because one of the elements in the mixing matrix is zero, the tri-bimaximal mixing does not allow CP violation to be manifest in neutrino oscillation, i.e. the Jarlskog parameter~\cite{jarlskog} $J$ is identically equal to zero. The combined experimental efforts~\cite{doubleCHOOZ,korean,japanese,daya,nove} will be able to measure CP violation in neutrino oscillation~\cite{lindner}.
Were CP violation to be
established in the future, it would definitively rule out tri-bimaximal mixing.  We also recall that CP violation in the lepton sector
has profound implication regarding why our universe is dominated by matter. There is thus an additional motivation to study deviations from  tri-bimaximal mixing.

Phenomenologically, small deviations from the tri-bimaximal pattern can be easily
parameterized in terms of three small parameters and studied~\cite{zee1,pakvasa}.
In \cite{hz}, we studied a minimal modification with one complex parameter. Here we extend this discussion.

Theoretically, many attempts have been made to derive the tri-bimaximal mixing, but in our opinion a simple and compelling construction is still sorely lacking. Many of the attempts were based on the tetrahedral group $A_{4}$ first studied by Ma and Rajasekaran for neutrino models~\cite{ma}, and subsequently by others~\cite{many}.  A group theoretic discussion was given in~\cite{Zee-physLett} attributing the difficulty to a clash between the residual $Z_{2}$ in one sector and $Z_{3}$ in another. Residual discrete symmetries in the context of neutrino mixing have also been extensively discussed in Ref.\cite{lam}. Authors in Ref.\cite{hkv} have named this the ``sequestering problem''.   Within the context of $A_{4}$, it was shown~\cite{kovtunzee} that two assumptions were needed to obtain an one-parameter family of mixing matrices which contains tri-bimaximal mixing. In other words, to obtain  tri-bimaximal mixing, it is necessary to find one reason or another to set this particular parameter to zero. This suggests, or at least motivates, studying various one-parameter modifications to
tri-bimaximal mixing.

\section{The form of minimal modifications for $V_{TB}$}

In our 2006 work~\cite{hz}, which we will review briefly in the appendices, we were naturally led in the context
of $A_{4}$, assuming that the sequestering problem of~\cite{hkv} could be solved, to a modification of  tri-bimaximal mixing in which the middle ``(1,1,1)'' column was left unchanged.

It has not escaped the notice of many authors that the 3 columns, $(-2,1,1),~(0,1,-1),~(1,1,1)$, of  $V_{TB}$
furnish the 2-dimensional and the 1-dimensional representations of the permutation group $S_{3}$. Historically, this observation led Wolfenstein\cite{wolf} long ago to guess, based on a sense that somehow $(1,1,1)$ is special,  a mixing matrix that consists of $V_{TB}$ with its last two columns interchanged.
Another early guess was put forward by Yamanaka, Sugawara, and Pakvasa\cite{Pak}. The mutual orthogonality of these 3 columns
of course also mean that they
correspond to the three Gell-Mann diagonal generators of SU(3). Similarly, these 3 columns
also appear~\cite{kovtunzee,bjorkenharrisonscott} in a table of Clebsch-Gordon coefficients for SU(2).
These may all indicate some deeper reasons for tri-bimaximal mixing, such as the possibility~\cite{kovtunzee} that neutrinos are composite.

With so little known about the underlying theory of neutrino mixing, we take here a purely phenomenological approach. If we take the column vectors as reflecting some
basic feature of neutrino, a minimal modification may be to keep the columns vectors as much as possible. Some phenomenological implications have been studies in Ref.\cite{red}.  In this paper we analyze these minimal modification to the tri-bimaximal mixing in light of the recent T2K data. Because of unitarity, we can leave at most one column unchanged. Minimal modification to the tri-bimaximal mixing can therefore be characterized by which column we leave unchanged. Some special cases have been considered in the literature~\cite{hz,kovtunzee,bjorkenharrisonscott,other,fl}. We refer to them as case  $V^a$, case  $V^b$, and case $V^c$, corresponding to keeping the third, second and first columns unchanged, respectively. This class of modifications can be obtained by multiplying a two generation mixing matrix from the right of $V_{TB}$, 2 and 3, 1 and 3, and 1 and 2 mixing patterns. These modifications can be viewed as perturbation to the tri-bimaximal by modifying the neutrino mass matrix~\cite{red}. One can also motivate minimal modification by perturbing the charged lepton mass matrix in a similar fashion which would result in one of the rows of $V_{TB}$ unchanged. One such an example has been studied by Friedberg and Lee in~\cite{fl}. Therefore there are another three types of minimal modification, keeping the first row or the second row or the third row indicated by case $W^a$, case $W^b$ and case $W^c$, respectively.

\subsection{One of the Columns in $V_{TB}$ unchanged }

Leaving one of the columns in $V_{TB}$ unchanged, we have the three possibilities:
\begin{eqnarray}
&&V^a = V_{TB} \left (\begin{array}{ccc}
\cos\tau&\sin\tau &0\\
-\sin\tau &\cos\tau &0\\
0&0&1
\end{array}
\right )\;, \;\;V^{b} = V_{TB} \left ( \begin{array}{ccc}
\cos\tau&0&\sin\tau  e^{i\delta}\\
0&1&0\\
-\sin\tau e^{-i\delta}&0&\cos\tau
\end{array}
\right )\;, \nonumber \\
&&V^c = V_{TB} \left (\begin{array}{ccc}
1&0&0\\
0&\cos\tau &\sin\tau e^{i\delta}\\
0&-\sin \tau e^{-i\delta}&\cos\tau
\end{array}
\right )\;. \label{vb}
\end{eqnarray}
We will use the abbreviation $c=\cos\tau$ and $s=\sin\tau$.

Note that for $V^a$, with the third column in $V_{TB}$ unchanged, $V_{e3}=0$ and there is no non-removable phase leading to a vanishing Jarlskog parameter, $J = 0$. No CP violation phenomena can show up in oscillation related processes.

\subsection{One Of The Rows In $V_{TB}$ unchanged}

Keeping the neutrino mass matrix unchanged and perturbing the charged lepton mass matrix for the tri-bimaximal mass matrices, one obtains a $V$ in the form
of a unitary $U$ multiplied from left to the tri-bimaximal mixing $V = U V_{TB}$. The three cases with one of the rows unchanged from tri-bimaximal mixing can be written in the following forms
\begin{eqnarray}
&&W^a = \left (\begin{array}{ccc}
1&0&0\\
0&c&s \\
0&-s&c
\end{array}
\right )V_{TB}\;,\;\;
W^b = \left ( \begin{array}{ccc}
c&0&s  e^{i\delta}\\
0&1&0\\
-s e^{-i\delta}&0&c
\end{array}
\right )V_{TB}\;, \nonumber\\
&&W^c = \left (\begin{array}{ccc}
c&s e^{i\delta} &0\\
-s e^{-i\delta}&c&0\\
0&0&1
\end{array}
\right )V_{TB}\;.
\end{eqnarray}
Since there is no non-removable phase in $W^a$, no CP violation phenomena can show up in oscillation related processes for this case.

As mentioned earlier, the case $W^{c}$ has been motivated theoretically and studied by Friedberg and Lee~\cite{fl}.

\section{Phenomenological Implications}

As was mentioned in the introduction, of these six cases, we are theoretically prejudiced in favor of $V^{b}$, which we studied in Ref.\cite{hz}. As far as we know, some of the other cases are not well motivated and we analyzed them merely as ``strawmen'' to be knocked down by future precision experiments.

There are no phases for cases $V^a$ and $W^a$. They are given by
\begin{eqnarray}
V^a = \left (\begin{array}{ccc}
{2c\over \sqrt{6}}  -{s\over \sqrt{3}}&{c\over \sqrt{3}} + {2s\over \sqrt{6}}&0\\
-{c\over \sqrt{6}} - {s\over \sqrt{3}}& {c\over \sqrt{3}} - {s\over \sqrt{6}}& {1\over \sqrt{2}}\\
-{c\over \sqrt{6}} - {s\over \sqrt{3}}&{c\over \sqrt{3}} - {s\over \sqrt{6}} &-{1\over \sqrt{2}}
\end{array}
\right )\;,\;\;
W^a = \left (\begin{array}{ccc}
{2\over \sqrt{6}}&{1\over \sqrt{3}}&0\\
-{c\over \sqrt{6}} - {s\over \sqrt{6}}& {c\over \sqrt{3}} + {s\over \sqrt{3}}& {c\over \sqrt{2}}-{s\over \sqrt{2}}\\
-{c\over \sqrt{6}} + {s\over \sqrt{6}}&{c\over \sqrt{3}} - {s\over \sqrt{3}} &-{c\over \sqrt{2}}+{s\over \sqrt{2}}
\end{array}
\right )\;,
\end{eqnarray}

The modification to tri-bimaximal is represented by a non-zero $s = \sin\tau$ ($c=\cos\tau$).
For case $V^a$, the main predictions of this mixing pattern are that $|V_{\mu 3}| = 1/\sqrt{2}$ which is at the boundary of the 1$\sigma$ allowed range. Present data constrain the modification mixing parameter $s$ (with $c>0$), which can modify the value for $V_{e2}$, to be in the range $-0.04 \sim 0.002 (-0.075 \sim 0.0)$ at 1$\sigma$ (3$\sigma)$ with the best fit value of -0.02. For case $W^a$, a definitive prediction for this case is that $V_{e2} = 1/\sqrt{3}$. This is outside the 1$\sigma$ range, but within 2$\sigma$ range. A non-zero $s$ will modify $V_{\mu3}$ away from $1/\sqrt{2}$. The current data allow $s$ to be in the range $0 \sim 0.08  (-0.14 \sim 0.105)$ at the 1$\sigma(3\sigma$) with the best fit value to be 0.08. For both cases $V^a$ and $W^a$, $V_{e3}$ are predicted to be zero. They are in conflict with data at the 3$\sigma$ level. Also there are no phases for $V^a$ and $W^a$. There is no CP violation in neutrino oscillation. This provides another test for cases $V^a$ and $W^a$. Should future experiments find CP violation in neutrino oscillations, these two cases would be ruled out.

The other four cases have two parameters, one can use available data on the magnitude of the elements in $V$ to constrain them and to predict other observables, in particular the Jarlskog CP violating parameter $J = Im(V_{e1}V_{e2}^*V_{\mu1}^*V_{\mu 2})$. Complete determination of parameters related to neutrino physics include the mixing angles and the CP violating Dirac phase, and also the absolute neutrino masses and possible Majorana phases.  Since not much information can be used to constrain the Majorana phases, in the following we will use  the combined pre-T2K and the recent T2K and MINOS data to study some phenomenological implications for the mixing parameters of the other four minimal modifications described in the previous section.
\\

\noindent
Case $V^b$

\bigskip
In this case the mixing matrix $V^b$ is given by
\begin{eqnarray}
V^b = \left (\begin{array}{ccc}
{2c\over \sqrt{6}}&{1\over \sqrt{3}}&{2 se^{i\delta}\over \sqrt{6}}\\
-{c\over \sqrt{6}} - {se^{-i\delta}\over \sqrt{2}}& {1\over \sqrt{3}}& {c\over \sqrt{2}}- {se^{i\delta}\over \sqrt{6}}\\
-{c\over \sqrt{6}} - {s e^{-i\delta}\over \sqrt{2}}&{1\over \sqrt{3}}&-{c\over \sqrt{2}}- {se^{i\delta}\over \sqrt{6}}
\end{array}
\right )\;.
\end{eqnarray}

A definitive prediction of this mixing pattern is that $V_{e2} = 1/\sqrt{3}$. This is outside the 1$\sigma$ range, but consistent with data within 2$\sigma$ level.
Since $V_{e2}$ is fixed to be
$1/\sqrt{3}$, $|V_{e3}|$ can be expressed as a function of $V_{e1}$ with
\begin{eqnarray}
V_{e3} = \sqrt{2/3 - |V_{e1}|^2}\;.
\end{eqnarray}
This can be used as an additional check.

In this case there is room for
a non-zero $V_{e3}$ and also a non-zero Jarlskog parameter given by $J = |V_{e1}||V_{e3}|\sin\delta/2\sqrt{3}$ for CP violation.
Using the allowed range for $\theta_{13}$, we can easily obtain information for $s$ with the best fit value given by 0.177 and the 1$\sigma$ (3$\sigma$) allowed range
$0.16\sim 0.22 (0.09 \sim 0.27)$.

Expressing $V_{e3}$ and $V_{\mu 3}$ in terms of the mixing angle $\tau$ (through $c$ and $s$) and CP violating phase, we have
\begin{eqnarray}
&&V_{e3}=\sqrt{\frac{2}{3}} s e^{i\delta}\;,\;\;V_{\mu 3}=\frac{1}{\sqrt{2}} c-\frac{1}{\sqrt{6} }s e^{i\delta}\;.\label{ddd}
\end{eqnarray}

One could use eq.(\ref{ddd}) to eliminate $\tau$ and relate $|V_{e3}|$,
$|V_{\mu 3}|$ and $\delta$,
\begin{eqnarray}
|V_{\mu 3}| &=& \left (\frac{1}{{2}} c^{2}+\frac{1}{{6} }s^{2} -\frac{1}{\sqrt{3} }c s \cos {\delta}\right )^{1/2}\nonumber\\
&=&\left [\left ({1\over 2}|V_{e3}|\cos\delta - {1\over \sqrt{2}}\sqrt{1- 3|V_{e3}|^2/2}\right )^2 + {1\over 4} |V_{e3}|^2\sin^2\delta \right ]^{1/2}\;.
\end{eqnarray}

\begin{figure}[h!]
\includegraphics[width=2.5in]{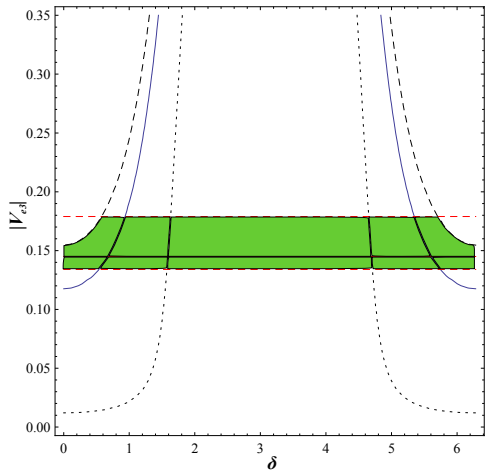}
\caption{Case $V^b$. $|V_{e3}|$ as a function of $\delta$ for $|V_{\mu 3}|$ equals to 0.617 (1$\sigma$ lower bound, dashed), 0.641 (best fit value, solid) and 0.701 (1$\sigma$ upper bound, dotted). The solid and dashed horizontal lines are for the best value and the 1$\sigma$ bounds of $|V_{e3}|$, respectively.
\label{fig1}}
\end{figure}

Using the known ranges of $\sin\theta_{23}$ and $\sin\theta_{13}$,  the 1$\sigma$ range for $V_{\mu 3}$ is determined to be in the range of $0.617 \sim 0.701$. We could plot
$|V_{e3}|$ in terms of the unknown $\delta$ for some typical values of
$V_{\mu 3}$ in the allowed range. In solving for $|V_{e3}|$ we have to pick the branch consistent with eq.(\ref{ddd}) of course. We will work in the convention where $s$ and $c$ are all positive. The result is shown in Fig.\ref{fig1}.
$|V_{e3}|$ is symmetric in the region of 0 to $\pi$ and $\pi$ to $2\pi$ for $\delta$. The allowed ranges in Table \ref{fit} rule out some portion of allowed $\delta$. Regions of $\delta$ close to $\pi$ are not allowed at the 1$\sigma$ level, but there are ranges of $\delta$ which can be consistent with data. Improved data can further narrow down the allowed range.

Since in this case there are only two unknown parameters, $s$ and $\delta$, in the model, the four parameters $\theta_{12}$, $\theta_{23}$, $\theta_{13}$ and the CP violating
parameter $\delta_{PDG}$ of the general parametrization are related. The boundaries in Fig. \ref{fig1} represent the 1$\sigma$ allowed ranges for the parameters of the model. The situations are similar for the other 3 cases in our later discussions. Note also that the analysis we are carrying out is insensitive to whether the neutrino mass hierarchy is normal or inverted.

\begin{figure}[h!]
\includegraphics[width=3in]{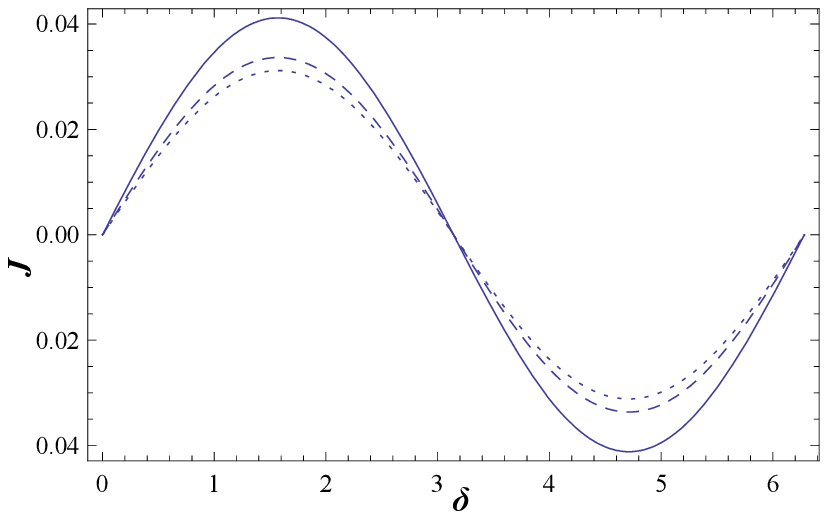}\hspace{1cm}\includegraphics[width=3in]{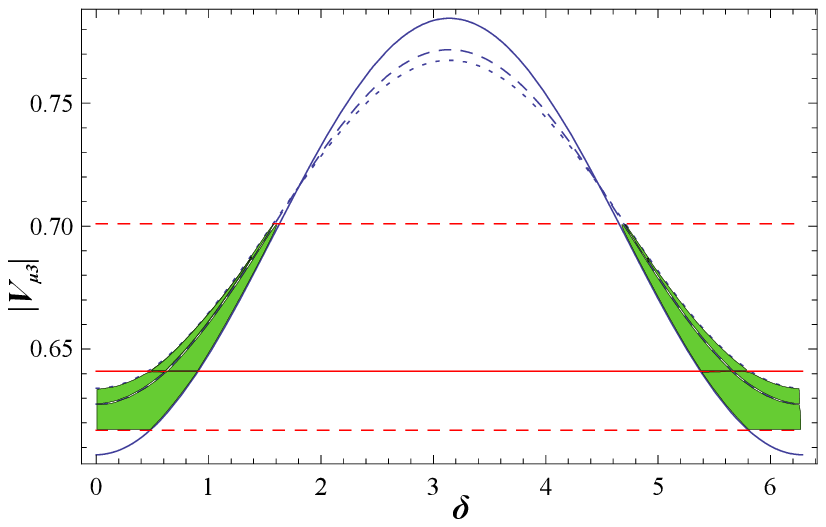}
\caption{Case $V^b$. Figure on the left: $J$ as a function of $\delta$ for $|V_{e3}|$ equals to  $0.179$ (solid), $0.145$ (dashed), and $0.134$ (dotted). Figure on the right: $|V_{\mu 3}|$ as a function of $\delta$ for $|V_{e3}|$ equals to  $0.179$ (solid), $0.145$ (dashed), and $0.134$ (dotted). The  solid and two dashed horizontal lines are for the best value and  the $1\sigma$ bounds of $|V_{\mu 3}|$, respectively.
\label{fig2}}
\end{figure}

\begin{figure}[h!]
\includegraphics[width=3in]{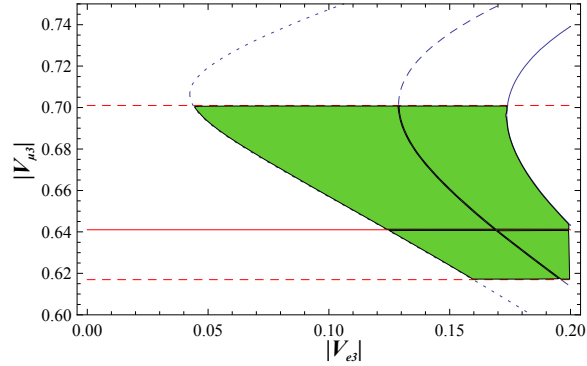}
\caption{Case $V^b$. Contours of $|V_{e3}|$ and $|V_{\mu 3}|$ for different values of $J$. The curves are for $J$ equals to  $\pm 0.04$ (solid), $\pm 0.03$ (dashed)  and $\pm 0.01$ (dotted). The  solid and two dashed horizontal lines are for the best value and  the $1\sigma$ bounds of $|V_{\mu 3}|$, respectively.
\label{fig3}}
\end{figure}

One can extract useful information on the CP violating parameter $J$ by using the relation
\begin{eqnarray}
J = {1\over 3 \sqrt{2}}\sin\delta |V_{e3}|\sqrt{1-3|V_{e3}|^2/2}\;.
\end{eqnarray}
Note that $J$ is simply proportional to $\sin\delta$. The results are shown in Fig.\ref{fig2}. On the left of Fig.\ref{fig2}, $J$ is plotted as a function of $\delta$ for three values of $|V_{e3}|$. The absolute value of $J$ can be as large as 0.04 for $|V_{e3}|$ takes its 1$\sigma$ upper value of $0.179$.

For $|V_{\mu 3}|$ close to its lower bound, $\delta$ close to 0 and $2\pi$ are favored. For $\delta$ close to $\pi$, $|V_{\mu 3}|$ is outside of its 1$\sigma$ allowed range. There are overlaps for the regions allowed in the right
figure of Fig.\ref{fig2} and those in Fig.\ref{fig1}.  Combining information from $V_{e2}$ discussed earlier, we can conclude that this case is ruled out at 1$\sigma$ level, but is consistent with data at 2$\sigma$ level.

In Fig.\ref{fig3},
the contours of $|V_{e3}|$ and $|V_{\mu 3}|$ for different values of $J$ are shown. The contours are degenerate in $\pm |J|$.
\\

\noindent
Case $V^c$

\bigskip
For this case, the mixing matrix is
\begin{eqnarray}
V^c = \left (\begin{array}{ccc}
{2\over \sqrt{6}}&{c\over \sqrt{3}}&{se^{i\delta}\over \sqrt{3}}\\
-{1\over \sqrt{6}}& {c\over \sqrt{3}}-{se^{-i\delta}\over \sqrt{2}}& {c\over \sqrt{2}}- {se^{i\delta}\over \sqrt{3}}\\
-{1\over \sqrt{6}}&{c\over \sqrt{3}}+{se^{-i\delta}\over \sqrt{2}}&-{c\over \sqrt{2}}+ {se^{i\delta}\over \sqrt{3}}
\end{array}
\right )\;.
\end{eqnarray}

A prediction for the mixing is that $V_{e1} = 2/\sqrt{6}$. $V_{e3}$ is related to $V_{e2}$ by \beq
|V_{e3}| = \sqrt{1/3 - |V_{e2}|^2}.\eeq Imposing unitarity of the $V_{PMNS}$, this is prediction in the $1\sigma$ region.

Within the $1\sigma$ range of $V_{e3}$, $s$ is allowed to vary from 0.2 to 0.31 with the best fit value of 0.25, and $V_{e2}$ can be within its 1$\sigma$ range.
One can again combine information from $|V_{\mu 3}|$ and $|V_{e3}|$ to constrain the CP violating phase $\delta$ and the Jarlskog parameter $J$. We have
\begin{eqnarray}
&&|V_{\mu 3} | = \left [\left (|V_{e3}|\cos\delta + {1\over \sqrt{2}} \sqrt{1-3|V_{e3}|^2} \right )^2 | + |V_{e3}|^2 \sin\delta^2\right ]^{1/2}\;,\nonumber\\
&&J = - {1\over 3\sqrt{2}}\sin\delta |V_{e3}| \sqrt{1-3|V_{e3}|^2} \;.
\end{eqnarray}

\begin{figure}[h!]
\includegraphics[width=2.5in]{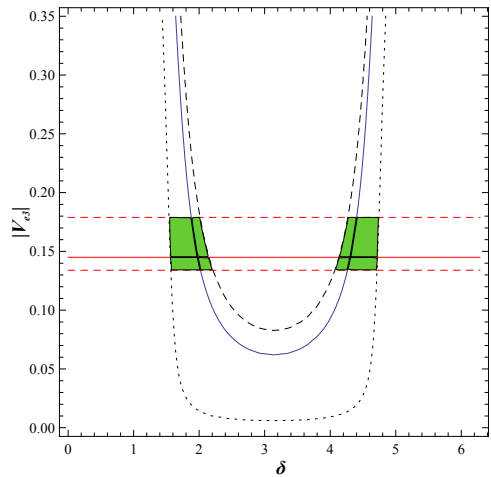}
\caption{Case $V^c$. $|V_{e3}|$ as a function of $\delta$ for $|V_{\mu 3}|$ equals to 0.617 (dashed), 0.641 (solid) and 0.701 (dotted). The solid and dashed horizontal lines are for the best value and the 1$\sigma$ bounds of $|V_{e3}|$, respectively.
\label{fig4}}
\end{figure}

\begin{figure}[h!]
\includegraphics[width=3in]{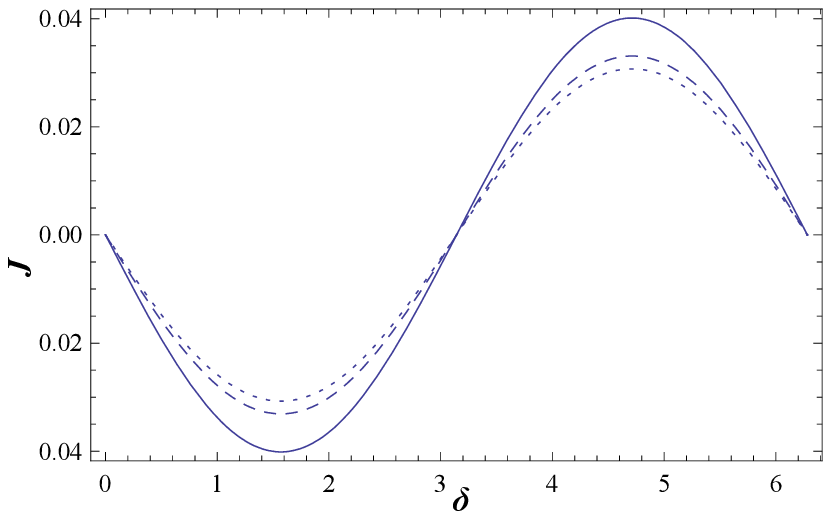}\hspace{1cm}\includegraphics[width=3in]{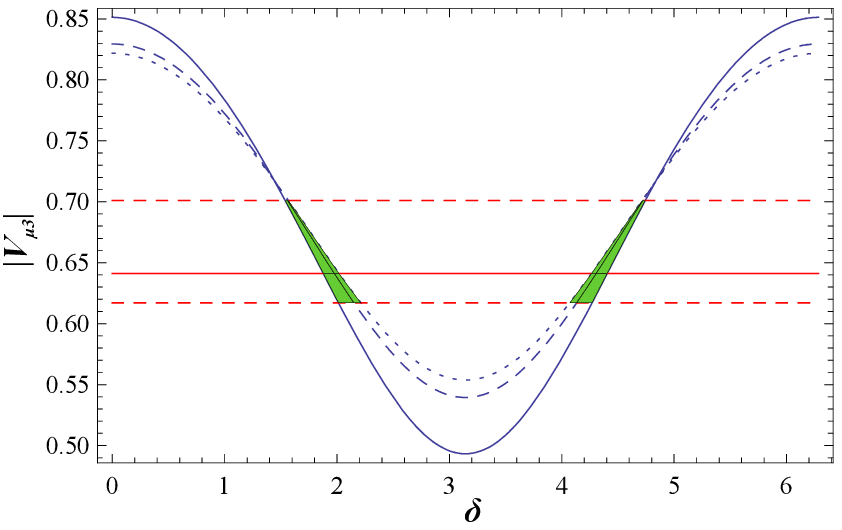}
\caption{Case $V^c$. Figure on the left: $J$ as a function of $\delta$ for $|V_{e3}|$ equals to  $0.179$ (solid line), $0.145$ (dashed line), and $0.134$ (dotted line) for case $V^c$. Figure on the right: $|V_{\mu 3}|$ as a function of $\delta$ for $|V_{e3}|$ equals to  $0.179$ (solid), $0.145$ (dashed), and $0.134$ (dotted). The solid and dashed horizontal lines are for the best value and the 1$\sigma$ bounds of $|V_{\mu 3}|$, respectively.
\label{fig5}}
\end{figure}

\begin{figure}[h!]
\includegraphics[width=3in]{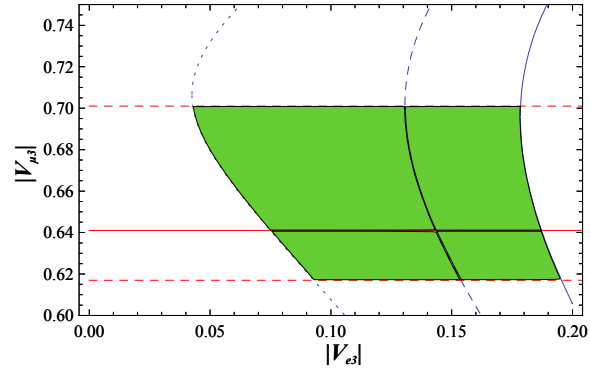}
\caption{Case $V^c$. Contours of $|V_{e3}|$ and $|V_{\mu 3}|$ for different values of $J$ (right). The curves are for $J$ equals to  $\pm 0.04$ (solid line), $\pm 0.03$ (dashed line)  and $\pm 0.01$ (dotted line). The solid and dashed horizontal lines are for the best value and the $1\sigma$ bounds of $|V_{\mu 3}|$, respectively.
\label{fig6}}
\end{figure}

The results are shown in Figs. \ref{fig4}, \ref{fig5} and \ref{fig6}. For this case even a larger portion of $\delta$ range is ruled out by data as can seen from Fig.{\ref{fig4}. From the right figure of Fig.\ref{fig5}, we also see that a large portion of $\delta$ is ruled out.  However, there are overlaps for the two regions. This case can be consistent with data at the 1$\sigma$ level.
CP violating information is shown on the left in Fig.\ref{fig5}. The largest value of $J$ is 0.04 for $|V_{e3}|$ equals to its 1$\sigma$ upper bound of $0.179$. The correlations of $J$, $V_{\mu3}$ and $V_{e3}$ are shown in Fig.\ref{fig6}.
\\

\noindent
Case $W^b$

\bigskip
The mixing matrix $W^b$ is given by
\begin{eqnarray}
W^b = \left (\begin{array}{ccc}
{2c\over \sqrt{6}}-{se^{i\delta}\over \sqrt{6}}&{c\over \sqrt{3}}+{se^{i\delta}\over \sqrt{3}}&-{se^{i\delta}\over \sqrt{2}}\\
-{1\over \sqrt{6}}& {1\over \sqrt{3}}& {1\over \sqrt{2}}\\
-{c\over \sqrt{6}}-{2s e^{-i\delta}\over \sqrt{6}}&{c\over \sqrt{3}}-{se^{-i\delta}\over \sqrt{3}}&-{c\over \sqrt{2}}
\end{array}
\right )\;.
\end{eqnarray}

This case predicts $V_{\mu 3} = 1/\sqrt{2}$. This prediction is at the boundary of 1$\sigma$ allowed range. But it is different than case $V^a$ since $V_{e3}$ is not zero and $J$ can be non-zero. $s$ is in the range of $0.19 \sim 0.25$ at 1$\sigma$ level with the best fit value given by 0.20. In this case correlations of $|V_{e2}|$, $|V_{e3}|$ and $\delta$ or $J$ are given by
\begin{eqnarray}
&&|V_{e2}| = {1\over \sqrt{3}} \left [\left( \sqrt{2}|V_{e3}|\cos\delta  + \sqrt{1-2|V_{e3}|^2}\right)^2 + 2|V_{e3}|^2\sin^2\delta\right ] ^{1/2}\;,\nonumber\\
&&J = {1\over 3\sqrt{2}}\sin\delta |V_{e3}|\sqrt{1-2|V_{e3}|^2}\;. \label{cor}
\end{eqnarray}

The results are shown in Figs.\ref{fig7}, ref{fig8} and \ref{fig9}. Now $|V_{e2}|$ is playing the role of $|V_{\mu 3}|$ for the cases $V^b$ and $V^c$.
From Fig.\ref{fig7} and the right figure in Fig\ref{fig8},
we see that the allowed range for $\delta$ is constrained to be even more closer to $\pi/2$ and $3\pi/2$. But there are still regions consistent with data at the slightly larger than
1$\sigma$ level.
$J$ can be as large as 0.04 for $|V_{e3}|$ taking its 1$\sigma$ upper of $0.179$. The correlations of $J$, $V_{\mu3}$ and $V_{e3}$ are shown in Fig.\ref{fig9}.
\\

\begin{figure}[h!]
\includegraphics[width=2.5in]{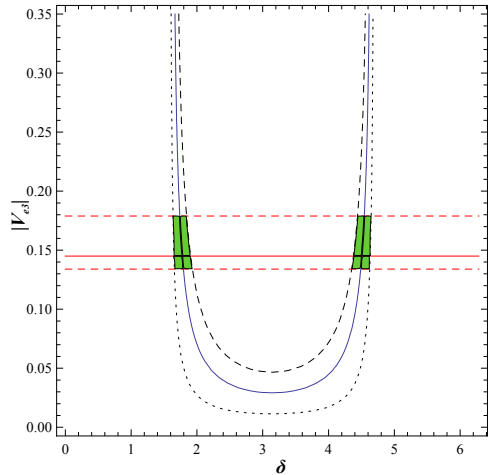}
\caption{Case $W^b$. $|V_{e3}|$ as a function of $\delta$ for $|V_{e2}|$ equals to 0.538 (dashed), 0.553 (solid) and 0.568 (dotted). The solid and dashed horizontal lines are for the best value and 1$\sigma$ bounds of $|V_{e3}|$, respectively.
\label{fig7}}
\end{figure}

\begin{figure}[h!]
\includegraphics[width=3in]{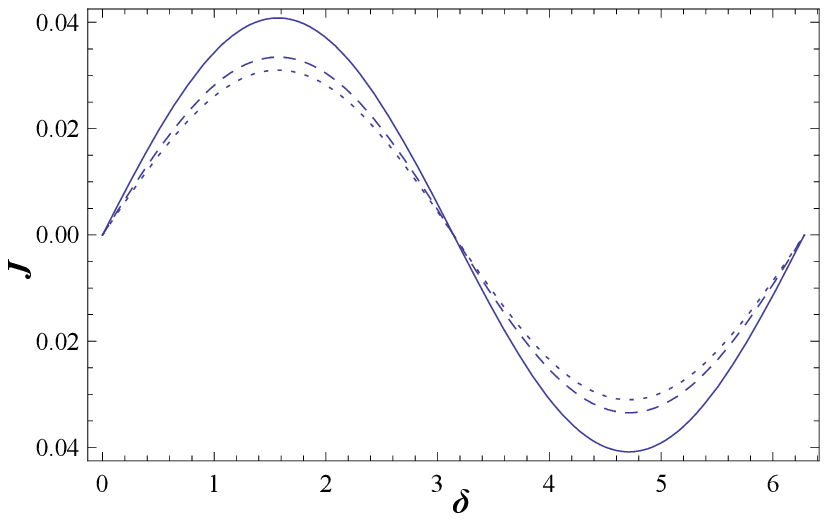}\hspace{1cm}\includegraphics[width=3in]{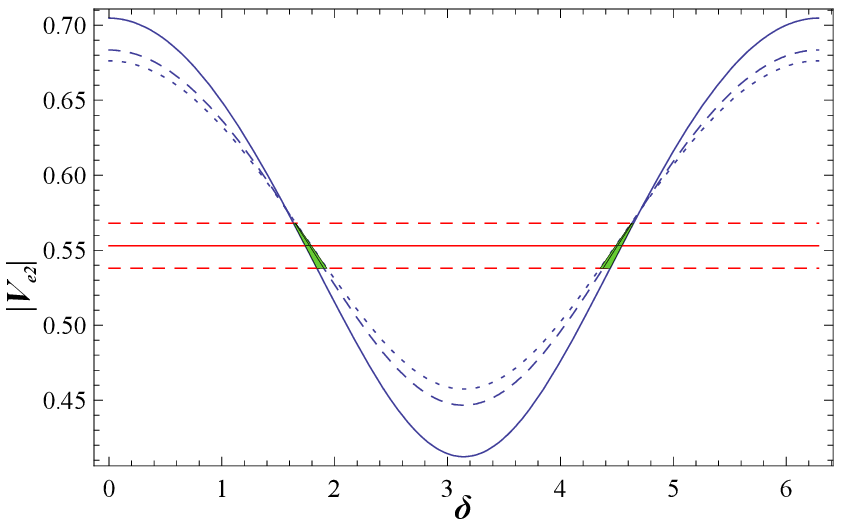}
\caption{Case $W^b$. Figure on the left: $J$ as a function of $\delta$ for $|V_{e3}|$ equals to  $0.179$ (solid), $0.145$ (dashed), and $0.134$ (dotted).  Figure on the right: $|V_{e2}|$ as a function of $\delta$ for $|V_{e3}|$ equals to  $0.179$ (solid), $0.145$ (dashed), and $0.134$ (dotted). The solid and dashed horizontal lines are for the best value and the 1$\sigma$ bounds of $|V_{e2}|$, respectively.
\label{fig8}}
\end{figure}

\begin{figure}[h!]
\includegraphics[width=3in]{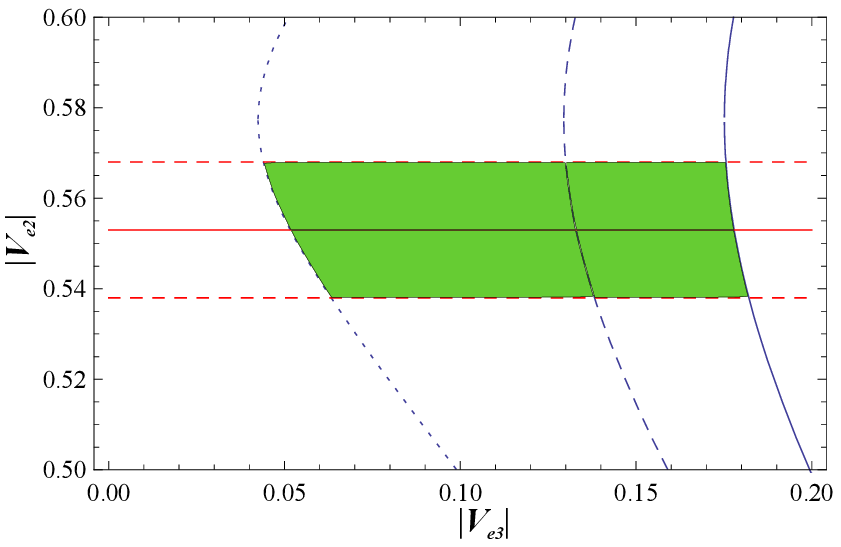}
\caption{Case $W^b$.  Contours of $|V_{e3}|$ and $|V_{e2}|$ for different values of $J$. The curves are for $J$ equals to  $\pm 0.04$ (solid), $\pm 0.03$ (dashed)  and $\pm 0.01$ (dotted).
The solid and dashed horizontal lines are for the best value and the 1$\sigma$ bounds of $|V_{e2}|$, respectively.
\label{fig9}}
\end{figure}

\noindent
Case $W^c$

 \bigskip In this case the third row is left unchanged from tri-bimaximal mixing~\cite{fl}and the mixing matrix $W^c$ is given by
\begin{eqnarray}
W^c = \left (\begin{array}{ccc}
{2c\over \sqrt{6}}- {s e^{i\delta}\over \sqrt{6}}&{c\over \sqrt{3}}+ {s e^{i\delta}\over \sqrt{3}}&{se^{i\delta}\over \sqrt{2}}\\
-{c\over \sqrt{6}}-{2se^{-i\delta}\over \sqrt{6}}& {c\over \sqrt{3}}-{se^{-i\delta}\over \sqrt{3}}& {c\over \sqrt{2}}\\
-{1\over \sqrt{6}}&{1\over \sqrt{3}}&-{1\over \sqrt{2}}
\end{array}
\right )\;.
\end{eqnarray}

We have
\begin{eqnarray}
&&|V_{e2}| = {1\over \sqrt{3}} \left [\left( \sqrt{2}|V_{e3}|\cos\delta  + \sqrt{1-2|V_{e3}|^2}\right)^2 + 2|V_{e3}|^2\sin^2\delta\right ] ^{1/2}\;,\nonumber\\
&&J = - {1\over 3\sqrt{2}}\sin\delta |V_{e3}|\sqrt{1-2|V_{e3}|^2}\;. \label{corr}
\end{eqnarray}

The expression for $|V_{e2}|$ is the same as for case $W^b$, but the sign of $J$ is minus of those for case $W^b$ when reading Figs.\ref{fig8} and \ref{fig9}, respectively. One can easily read off the constraints on $\delta$ and $J$ from Figs.\ref{fig7} and \ref{fig8}. {\bf A crucial difference is that $V_{\mu 3}$ is no longer $1/\sqrt{2}$, but $c/\sqrt{2}$. $|V_{\mu 3}|$ can vary from $0.684 \sim 0.694 (0.670 \sim 0.703)$ at the 1$\sigma$ (3$\sigma$) level. This case is consistent with data at 1$\sigma$ level.} Precise measurement of $|V_{\mu 3}|$ can be used to distinguish these two cases.

\section{Summary}

Recent data from T2K and MINOS show evidence of a non-zero $V_{e3}$ at 90\% C.L. level. There may be the need for modifications to the tri-bimaximal mixing with which $V_{e3}$ is equal to zero resulting in a CP conserving mixing. We have studied several possible ways to minimally modify the tri-bimaximal mixing by keeping one of the columns or one of the rows in the tri-bimaximal mixing unchanged. Six cases were studied. Two of the cases have $V_{e3} = 0$. These two cases are in tension with data at 3$\sigma$ level. Also for these two cases, the CP violating Jarlskog parameter $J$ is identically zero. CP violation in neutrino oscillation can provide new test for these two cases. For the other four cases, all have two parameters in the mixing matrix. Current data on neutrino oscillation can put constraints on the parameters, but consistent with data within 2$\sigma$ for case $V^b$, and within 1$\sigma$ for the other 3 cases. The allowed ranges for CP violation are also constrained. Future experiments can test the predictions to rule out these models.

\bigskip
\noindent
{\bf Acknowledgment}: We thank Koji Tsumura for technical assistance with figures.
This work was carried while AZ was at the Academia Sinica, whose hospitality is gratefully acknowledged where this work was initiated in the summer of 2010.
This work was partially supported by NSF under Grant No. 04-56556, NSC, NCTS and SJTU 985 grant.

\appendix

\section{}

Here we review some features of the family group $A_4$~\cite{ma} that led us to favor $V^b$, which we analyzed in~\cite{hz}. A key point to achieve the tri-bimaximal mixing in models based on $A_4$ symmetry is to obtain the matrices $U_l$ and $U_\nu$ which diagonalize the charged lepton and neutrino mass matrices $M_l$ and $M_\nu$, $U^\dagger_l M_l U_r = D_l$ and $U^T_\nu M_\nu U_\nu = D_\nu$ in the following forms~\cite{hz}
\begin{eqnarray}
U_l = {1\over \sqrt{3}}\left (\begin{array} {ccc}
1&\omega^2&\omega\\ 1&1&1\\1&\omega&\omega^2 \end{array} \right )\;,\;\;U_\nu = {1\over \sqrt{2}}\left (\begin{array} {ccc}
1&0&-1\\ 0&\sqrt{2}&0\\1&0&1 \end{array} \right )\;,\label{rightform}
\end{eqnarray}
where $\omega^3 = 1$. The mixing matrix $V$ is given by $U^\dagger_l U_\nu$. Note that $U_r$ which plays a role in diagonalizing the charged lepton mass does not show up in $V$. With suitable choices of phase conventions, $V$ can be written in the form in eq.\ref{vtri}. In general the elements in the diagonal matrix $D_\nu$ have phases, the Majorana phases. We provide some details in Appendix B for obtaining the relevant mass matrix.

One can easily find that the neutrino mass matrix must be the following form
\begin{eqnarray}
M_\nu = \left ( \begin{array}{ccc} \alpha&0&\beta\\0& \gamma&0\\\beta&0&\alpha\end{array} \right ),
\end{eqnarray}
to obtain the right form for $U_\nu$.
With appropriate redefinition of phases and mixing angles, one obtained the form $V_{TB}$ for $V$.

To achieve the above with specific models, there are some requirements for Higgs boson fields vacuum expectations values as can be seen in Appendix B. In general, $M_\nu$ above will be modified. If one keeps the charged lepton mass matrix unchanged, the most general Higgs potential may not respect the conditions leading to modification resulting modifying the `11' entry to $\alpha-\epsilon$ and the `33' entry to $\alpha + \epsilon$ for the neutrino mass matrix $M_\nu$ given by eq.\ref{appen-1} in Appendix B. With appropriated redefinition of phases and angles shown in Appendix B, one obtains
$V^b$
where we have redefined $\delta = \eta + \pi/2$ to the form in eq.\ref{appen-2}.

Note that in taking the form of $V^b$ in eq.\ref{vb}, without Majorana phases in $V$, we have absorbed possible Majorana phases in the masses $\tilde m_i = m_i e^{i2\kappa_i}$. Here $m_i$ are real and positive. Without loss of generality, one can always choose one of the $\kappa_i$ to be zero, for example $\kappa_2 =0$.
The above form belongs to the minimal modification to the $V_{TB}$ mixing pattern specified earlier with the elements in the second column remain to be $V_{\alpha 2 } = 1/\sqrt{3}$.

We have arrived at this minimal modification from a specific model. In fact, the above parametrization is the most general one for $V_{\alpha 2} = 1/\sqrt{3}$ up to phase conventions. One can understand this by starting with a most general parametrization used by the Particle Data Group, $V_{PDG}$ and then set certain angles to some particular values to make sure that $V_{\alpha 2} = 1/\sqrt{3}$. Since we want all of the
elements in the second column to be $1/\sqrt{3}$, for convenience we exchange the second and the third columns, and move the Dirac phase at different locations, by redefining the phase of charged leptons and neutrinos, according to the following
\begin{eqnarray}
V^b = P_L V_{PDG} E P_R,
\end{eqnarray}
where $P_{L,R}$ diagonal phase matrices with elements, $P_L = diag(e^{i\delta}, 1, 1)$, $P_R = diag(e^{-i\delta}, 1, 1)$ and
$E$ is a matrix switching the second and third columns with elements $E_{ij} = \delta_{i1}\delta_{j1} + \delta_{i2}\delta_{j3} + \delta_{i3}\delta_{j2}$.

We have
\begin{eqnarray}
V^b = \left (\begin{array}{lll}
c_{13}c_{12}& s_{13}& c_{13}s_{12} e^{i\delta}\\
-s_{12}c_{23}e^{- i\delta} - c_{12}s_{23}s_{13}& s_{23}c_{13}& c_{12}c_{23} - s_{12}s_{23}s_{13}e^{i\delta}\\
s_{12}s_{23}e^{-i\delta} - c_{12}c_{23}s_{13}& c_{23}c_{13}& -c_{12}s_{23} - s_{12}c_{23}s_{13}e^{i\delta}
\end{array} \right )\;.
\end{eqnarray}
To have all of the elements in the second column in the above matrix to be $1/\sqrt{3}$, one just needs to set $s_{13} =1/\sqrt{3}$, $c_{13} = \sqrt{2/3}$ and $s_{23} = c_{23} = 1/\sqrt{2}$. We then obtain eq.\ref{vb} with $c = c_{12}$ and $s = s_{12}$.

In this basis, one can reconstruct the elements in the neutrino mass matrix in terms of the mixing angle $\tau$, phase $\delta$ and masses $\tilde m_i$ by requiring $M_\nu = U^*_lV D_\nu V^T U^\dagger_l$. We have
\begin{eqnarray}
&&\alpha = {1\over 2} \left ( (c^2 - s^2 e^{-2i\delta}) \tilde m_1 - (c^2 - s^2 e^{2i\delta}) \tilde m_3\right )\;,\nonumber\\
&&\beta = {1\over 2} \left ((c^2 + s^2 e^{-2i\delta}) \tilde m_1 + (c^2 + s^2 e^{2i\delta}) \tilde m_3\right )\;,\nonumber\\
&&\epsilon = i cs (e^{-i\delta} \tilde m_1 - e^{i\delta} \tilde m_3)\;,\;\;\gamma = \tilde m_2.
\end{eqnarray}
The $\alpha$, $\beta$ and $\epsilon$ here are equivalent to those given in Appendix B with a different basis. The important thing is that the  number of independent parameters are the same, total six of them.

\bigskip

\section{}

In Appendix B, we briefly outline how $V^b$ can be obtained from models based on $A_4$ family symmetry\cite{ma} with the standard model (SM) gauge symmetry.

A way to obtain $U_l$ in eq.\ref{rightform} is to assign the three left handed SM lepton doublet $l_L = (l_{L1},\;l_{L2},\;l_{L3})$, and three right handed neutrino $(l_{R1},\;l_{R2},\;l_{R3})$ representations into a triplet, and  the three singlets $1$, $1'$ and $1''$ of $A_4$ group, respectively. A SM doublet and $A_4$ triplet Higgs representation $\Phi = (\Phi_1,\;\Phi_2,\;\Phi_3)$ then leads to the Yukawa coupling terms
\begin{eqnarray}
L_l = \lambda_e (\bar l_L \Phi_1) l_{R1} + \lambda_\mu \omega^2 (\bar l_L \Phi)_{1'}l_{R2} + \lambda_\tau \omega (\bar l_L \Phi)_{1''} l_{R3} + h.c.\;.\label{charge}
\end{eqnarray}
After $\Phi$ develops a vev of the form $<\Phi> = (v_\Phi,v_\Phi,v_\Phi)$, the charged lepton mass matrix is given by, $M_l = U_l D_l$ with $D_l = Diag(m_e,\;m_\mu,\;m_\tau)$ and $m_i = \sqrt{3}v_\Phi\lambda_i$. This gives the right $U_l$ with $U_r = I$.

To obtain the right form of neutrino mass matrix, three right handed neutrinos $\nu_R = (\nu_{R1},\;\nu_{R2},\;\nu_{R3})$ of $A_4$ triplet, a SM doublet and $A_4$ singlet Higgs $\phi$, and a SM singlet and $A_4$ triplet Higgs $\chi = (\chi_1,\;\chi_2,\;\chi_3)$, are needed. The Yukawa coupling terms relevant are
\begin{eqnarray}
L_\nu = \lambda_\nu (\bar l_L \nu_R)_1 + m (\bar \nu_R \nu^C_R)_1 + \lambda_\chi ((\nu_R\nu^C_R)_3 \chi)_1 + h.c.
\end{eqnarray}
With the vev structure $<\phi> = v_\phi$ and $<\chi> = (0,v_\chi, 0)$, the neutrino mass matrix is given by
\begin{eqnarray}
M = \left ( \begin{array}{cc} 0&M_D\\M_D^T& M_R \end{array} \right ),\;\;M_R = \left ( \begin{array}{ccc} m&0&m_\chi\\0&m&0\\m_\chi&0&m \end{array} \right )\;,
\end{eqnarray}
where $M_D = Diag(1,1,1) \lambda_\nu v_\phi$, and $m_\chi = \lambda_\chi v_\chi$. The light neutrino mass matrix is $M_\nu = - M_D M_R^{-1}M^T_D$ which is the desired form giving $U_\nu$ in eq.\ref{rightform}.

The charged lepton masses are controlled by the vev of $\Phi$ which requires the components of $\Phi$ to have the same vev which leave  $Z_3$ unbroken symmetry in the theory, while the neutrino mass matrix is controlled by the vev of $\phi$ and $\chi$ with only $<\chi_2>$ non-zero preserving a $Z_2$ unbroken symmetry. If there is no communication between these two sectors, the residual $Z_3$ and $Z_2$ symmetries are left unbroken. But in general these two sectors cannot be completely sequestered and can interact which complicates the situation\cite{Zee-physLett,hkv}, {\bf for example the vev structure with only $<\chi_2>$ be non-zero may not be maintained.} One of the consequences concern us is that the mass matrix $M_\nu$ will be modified to\cite{hz}
\begin{eqnarray}
M_\nu = \left ( \begin{array}{ccc} \alpha -\epsilon&0&\beta\\0& \gamma&0\\\beta&0&\alpha +\epsilon \end{array} \right ).\label{appen-1}
\end{eqnarray}
The above will lead to a different form for $U_\nu$ from that in eq.\ref{rightform} which can be written as
\begin{eqnarray}
U_\nu = \left ( \begin{array}{ccc}
-c_\theta &0&-is_\theta\\
0&1&0\\
-s_\theta e^{i\rho}&0&ic_\theta e^{i\rho}
\end{array} \right )\;,
\end{eqnarray}
where $c_\theta=\cos\theta$ and $s_\theta=\sin\theta$ with
\begin{eqnarray}
&&\tan^2(2\theta) = {4|\beta|^2\over (|\alpha + \epsilon| - |\alpha - \epsilon|)^2} ( 1- {4|\alpha^2 - \epsilon^2|\over (|\alpha + \epsilon| + |\alpha - \epsilon|)^2} \sin^2\sigma)\;,\;\;\sigma = \arg({\beta\over \sqrt{\alpha^2-\epsilon^2}}) \;,\nonumber\\
&&\rho = \tilde \delta +\arg(\sqrt{{\alpha - \epsilon\over \alpha +\epsilon}}) \;,\;\;\tan\tilde \delta = {|\alpha + \epsilon| - |\alpha -\epsilon|\over |\alpha + \epsilon| + |\alpha - \epsilon|}\tan\sigma\;.
\end{eqnarray}
The phase conventions are chosen such that setting $\epsilon = 0$, the resulting mixing matrix $V$ goes to $V_{TB}$.

In this basis, $V$ is given by
\begin{eqnarray}
V = {1\over \sqrt{3}}\left ( \begin{array}{ccc}
-(c_\theta+s_\theta e^{i\rho})&1&i (c_\theta e^{i\rho} - s_\theta)\\
-(\omega c_\theta +\omega^2 s_\theta e^{i\rho})&1&i(\omega^2 c_\theta e^{i\rho} - \omega s_\theta)\\
-(\omega^2 c_\theta + \omega s_\theta e^{i\rho})&1&i(\omega c_\theta e^{i\rho} - \omega^2 s_\theta)
\end{array}
\right )\;,
\end{eqnarray}
which can be further rewritten as
\begin{eqnarray}
V = V_{TB} \left ( \begin{array}{ccc}
\cos\tau &0&i \sin\tau e^{i\eta}\\
0&1&0\\
i\sin\tau e^{-i\eta}&0&\cos\tau
\end{array}
\right )V_p\;, \label{appen-2}
\end{eqnarray}
where $V_p$ is a diagonal phase matrix $V_p = (e^{i(\xi + \rho/2)}, 1, e^{i(-\xi + \rho/2})$ multiplied from right. $\tau$, $\eta$ and $\xi$ given by
\begin{eqnarray}
\sin^2\tau = {1\over 2} (1-\sin(2\theta)\cos\rho)\;,\;\;\tan\xi = -{1-\tan\theta \over 1+\tan\theta} \tan(\rho/2)\;,\;\;
\tan\eta = \tan(2\theta) \sin\rho\;.
\end{eqnarray}

Absorbing the phases in $V_p$ into the neutrino masses, the total Majorana phases for $\tilde m_1$, $\tilde m_2$ and $\tilde m_3$ are $\kappa_1=-(2\xi + \rho + 2\alpha_1)$, $\kappa_2=\alpha_2$ and $\kappa_3 = -(-2\xi + \rho + 2\alpha_3)$ with
\begin{eqnarray}
&&\alpha_1 = -{1\over 2} [\arg(c_\theta^2|\alpha-\epsilon| + 2s_\theta c_\theta|\beta|e^{i(\tilde \delta + \sigma)} + s_\theta^2 |\alpha +\epsilon|e^{2i\tilde \delta})] - \arg(-\sqrt{\alpha -\epsilon})\;,\nonumber\\
&&\alpha_2 = -{1\over 2} \arg(\gamma)\;,\nonumber\\
&&\alpha_3 = -{1\over 2} [\arg(c_\theta^2|\alpha-\epsilon| - 2s_\theta c_\theta|\beta|e^{i(\tilde \delta + \sigma)} + s_\theta^2 |\alpha +\epsilon|e^{2i\tilde \delta})] - \arg(i\sqrt{\alpha +\epsilon}) \;.
\end{eqnarray}
In this basis, $V$ is given by eq.\ref{appen-2}, but with $V_p$ removed.

The absolute masses squared are given by
\begin{eqnarray}
&&m_1^2 = |\alpha|^2 +|\epsilon|^2 +|\beta|^2 - {2Re(\alpha\epsilon^*)\over \cos(2\theta)}\;,\nonumber\\
&&m^2_2 = |\gamma|^2\;,\nonumber\\
&&m_3^2 = |\alpha|^2 +|\epsilon|^2 +|\beta|^2 + {2Re(\alpha\epsilon^*)\over \cos(2\theta)}\;.
\end{eqnarray}

\end{document}